\documentclass{acmhack}
\usepackage{xspace}
\usepackage{graphicx}
\usepackage{fancyvrb}
\usepackage{listings}
\fvset{fontsize=\footnotesize,frame=single}

\newdef{definition}[theorem]{Definition}
\newdef{remark}[theorem]{Remark}
\newcommand{\mspace}{\hspace*{0mm}}

\newcommand{\slang}{{\rm S-Lang}\xspace}
\newcommand{\Hypot}{{\tt hypot}\xspace}
\newcommand{\Sin}{{\tt sin}\xspace}
\newcommand{\Cos}{{\tt cos}\xspace}
\newcommand{\Log}{{\tt log}\xspace}
\newcommand{\Exp}{{\tt exp}\xspace}
\newcommand{\where}{{\tt where}\xspace}
\newcommand{\atof}{{\tt atof}\xspace}
\newcommand{\vmult}{{\tt vmult}\xspace}

\newcommand{\slirp}{{\sc SLIRP}\xspace}
\newcommand{\isis}{{\sc ISIS}\xspace}
\newcommand{\omp}{{\rm OpenMP}\xspace}
\newcommand{\lino}{{\rm Linux1}\xspace}
\newcommand{\lint}{{\rm Linux2}\xspace}
\newcommand{\solf}{{\rm Solaris4}\xspace}
\newcommand{\solo}{{\rm Solaris1}\xspace}
\newcommand{\M}{{\tt M}\xspace}

\newcommand{\A}{{\tt A}\xspace}
\newcommand{\E}{{\tt E}\xspace}
\newcommand{\D}{{\tt D}\xspace}

\title{Getting More From Your Multicore: Exploiting OpenMP From An Open Source Numerical Scripting Language}

\author{Michael S. Noble\\Kavli Institute for Astrophysics, Massachusetts Institute of Technology\\June 26, 2007}
            
\begin{abstract}
We introduce \slirp, a module generator for the \slang numerical scripting
language, with a focus on its vectorization capabilities. We demonstrate how
both \slirp and \slang were easily adapted to exploit the inherent parallelism
of high-level mathematical languages with \omp, allowing general users to
employ tightly-coupled multiprocessors in scriptable research calculations
while requiring no special knowledge of parallel programming.  Motivated by
examples in the \isis astrophysical modeling \& analysis tool, performance
figures are presented for several machine and compiler configurations,
demonstrating beneficial speedups for real-world operations.
\end{abstract}

\category{D.3.2}{Programming Languages}{Language Classifications}[Very high-level languages, Concurrent, distributed, and parallel languages]
\category{D.3.4}{Programming Languages}{Processors}[Code generation]
\category{D.2.8}{Software Engineering}{Metrics}[Performance measures]
\keywords{Scientific Computation, Interactive Analysis, Astrophysics}
\begin{document}

\maketitle

\section{Introduction}

We are witnessing the arrival of serious multiprocessing capability on
the desktop.  While single-user machines with multiple CPUs have been
available for several years, they remain uncommon and have typically only
doubled, or to a lesser extent quadrupled, the number of processors.
That is changing, though, as multicore chip designs begin to make it
economical for typical users to access 8, 16, or even more processor cores.
There is widespread concern, however, that it will not be easy to harness
all of this power {\em within} applications on the desktop
\cite{1095423,1168901}.
Most desktop software has been developed for 
one CPU, and writing multithreaded software has traditionally
been difficult \cite{10.1109/MC.2006.180}.

Drawing from our own scientific niche, we recently noted how
rarely parallel computing is employed for common modeling and analysis
computations in observational astrophysics \cite{2006ASPC..351..481N}.
Researchers in other fields indicate a similarly low adoption of parallel
methods by general investigators in their disciplines, e.g. \cite{1095677}.
Moreover, even if
parallel programming were ``easier," the large bodies of serial
software developed in communities over decades, and the mindsets they
embody, cannot be changed overnight.  Another difficulty is maintaining
trust in such codes -- instilled by years of vetting through the process
of scientific publication -- as they are retrofitted for parallelism.  It
has therefore been easier and safer for general practicioners to increase
performance by purchasing
faster serial hardware, rather than revamping algorithms or techniques
for parallelism.  Chip manufacturers are effectively telling us with
multicore designs that this tactic will not remain viable for much longer.

\subsection{Very High Level Numerical Languages}
At the same time researchers are well versed in scripting, particularly
with array-oriented numerical languages like MatLab, Octave,
and \slang, to name just a few.  A key feature of these languages
is that they allow easy manipulation of mathematical structures of
arbitrary dimension, combining the brevity and convenience of an
interpreted environment with most of the performance of compiled code.
Operators and functions defined in the language of implementation (e.g. C)
to work only with scalars are extended to interpreted arrays in the natural
way, facilitating concise expressions such as
{\tt c  = sin(a\^{}3) + b*10.0}
without regard to whether {\tt a} or {\tt b} are scalars, vectors,
or multidimensional arrays.  The high performance stems from moving array
traversals out of the interpreted layer and into lower-level code like
this fragment of C which provides vectorized multiplication in \slang:
\begin{Verbatim}[xleftmargin=4cm,frame=none]
case SLANG_TIMES:
  ...
  for (n = 0; n < na; n++)
      c[n] = a[n] * b[n];
  ...
\end{Verbatim}
\noindent
One of the earliest motivations for \slirp, the module generator
for \slang, was to extend the reach of this vectorization to external
C/C++ and Fortran codes.

\subsection{\omp}
The code above suggests that much of the strength and appeal of
numerical scripting languages stems from relatively simple internal loops
over regular structures.
Another advantage of having these regular loops in lower-level compiled
codes is that they are ripe for parallelization with \omp, a hardware-neutral
specification aimed at facilitating parallel programming on shared memory
multiprocessors.  Conformant implementations of \omp offer a set
of compiler directives for C/C++ or Fortran programs, supporting
libraries, and environment variables which tune their operation.  Programs
are parallelized with \omp by tagging regions of code with
comments in Fortran or preprocessor directives in C/C++.
Proponents contend that conceptual simplicity makes \omp more approachable
than other parallel programming models, e.g. message-passing in MPI or PVM,
and emphasize the added benefit of allowing single bodies of code to be used
for both serial and parallel execution.  For instance, changing the above
loop to
\begin{Verbatim}[xleftmargin=4cm,frame=none]
#pragma omp parallel for
for (n = 0; n < na; n++)
    c[n] = a[n] * b[n];
\end{Verbatim}
\label{openmp}
parallelizes the \slang multiplication operator; the
pragma is simply ignored by a non-conformant compiler, resulting in a
sequential program.  \omp
runtime environments have also been used for distributed computation over
networked clusters, but this is not yet within the standard.  Despite the
promise of straightforward parallelism, the spread of \omp beyond high
performance computing research groups has in part been hindered by the
need for special compiler support.  With a few exceptions such as
OdinMP \cite{OdinMP.Nov2004} and Omni \cite{690388}, this has come largely
in the form of commercial compilers.  The shortfall of free compiler 
support for \omp helps explain the relatively small presence of \omp in
open source numerical software, and marks the availability of \omp in
GCC as a significant step towards the wider adoption of parallel computing
by general practicioners.
\subsection{Testbed}
Our work was conducted primarily on 2 machine
configurations: a dual-CPU (1.8 Ghz) Athlon workstation with 2 GB RAM
running Debian 3.1 GNU/Linux, and 4 of 8 CPUs (750 Mhz) on a Solaris
5.9 server with 32 GB Ram.  We refer to these as \lint and \solf,
and use \lino and \solo to denote serial execution.  The author was
the only user of \lint for all tests, while \solf was shared with users
running various jobs, many compute-intensive.  Versions 1.9.3 and 2.0.7
of \slirp and \slang were used, 
with codes executed
in {\tt slsh} and version 1.4.7 of \isis \cite{2002hrxs.confE..17H}, an
astrophysical modeling and analysis tool developed at MIT.

\section{Related Work}

We now highlight some of the more prominent efforts which relate to 
the three main areas encompassed by this paper: high-level numerical
scripting environments, simplifying the use of tightly-coupled
multiprocessors within them, and wrapper generators for them.  In the
commercial arena MatLab and Mathematica are among the most widely
used high-level numerical environments to provide simplified support
for multiprocessing.
The MatLab*P extension strives to make it easy to adapt MatLab scripts for
parallelism, by transparently storing arrays on, and offloading computations
upon them to, a distributed compute server.  IDL is arguably the most popular
high level language in our astrophysics community; it also makes the use
of multiple CPUs transparent, through an internal thread pool.
The clear contrasts between these efforts and our work are cost and
openness: commercial packages, while at times preceding their open source
equivalents or eclipsing them in features, can be expensive to purchase and
maintain.  In addition, many research projects require software customizations
which might either violate the proprietary nature of closed source or introduce
unacceptable delays while waiting for fulfillment by vendors.  Such conditions
generally lead to the embrace of open methods.

In the realm of open software,
OctaveHPC\footnote{http://www.hpc.unsw.edu.au/OctaveHPC.html} was created
to generalize Octave, a MatLab clone, for 64-bit platforms and integrate
\omp directives so as to make
transparent use of multiprocessors.  The 64-bit extensions have been folded
in to the source repository, but as of version 2.9.10 no \omp constructs
appear within the Octave codebase nor do any papers appear to have been
published describing the work.
OdinMP \cite{OdinMP.Nov2004} and Omni \cite{690388} are notable for being
among the earliest open-source compilers supporting \omp.  Oriented towards
academic research, neither has penetrated the wider open source community to
nearly the same extent as GCC, nor do they appear to be actively maintained.
The cOMPunity group\footnote{http://www.compunity.org} maintains a list
of additional free \omp tools.  SWIG \cite{1996SWIG} is arguably the
most powerful and widely used wrapper generator in the world.
It can generate bindings to numerous scripting languages and has considerably
deeper support for C++ than does \slirp.  Unlike \slirp, SWIG will not wrap
Fortran, nor does it generate vectorized wrappers for numerical languages.
The MatWrap\footnote{http://freshmeat.net/projects/matwrap} tool has been
used to generate vectorized bindings to the MatLab, Tela, and Octave matrix
languages.  Vectorization in MatWrap is not as advanced as in \slirp (e.g.
no support for Fortran, strings, or parallelization; arrays of unlike
dimension may not be mixed), and it has not been maintained since 2001.
The Tcl, Perl, and Python scripting languages are not vectorized, but
actively-maintained numerical extensions do exist for each: BLT \& NAP for
TCL, PDL for Perl, and Numeric,
NumArray, \& NumPy for Python.  Although some work has been done to create a
SWIG extension which takes advantage of Python numerical extensions, it is
not clear from the literature that vector-parallel wrappers can be
auto-generated for these languages, particularly with the ease
of
\slirp.

\section{\slang}
\slang is an extensible, C-like scripting language used in a number of popular
free software packages and bundled with every major Linux distribution. It is
highly suitable for scientific and engineering computation, offering a wide
selection of extension modules and multidimensional numerics on par with
commercial packages like MatLab and IDL.  While a comprehensive analysis of
the
\begin{figure*}[p]
  \begin{centering}
  \includegraphics[angle=-90,scale=0.28]{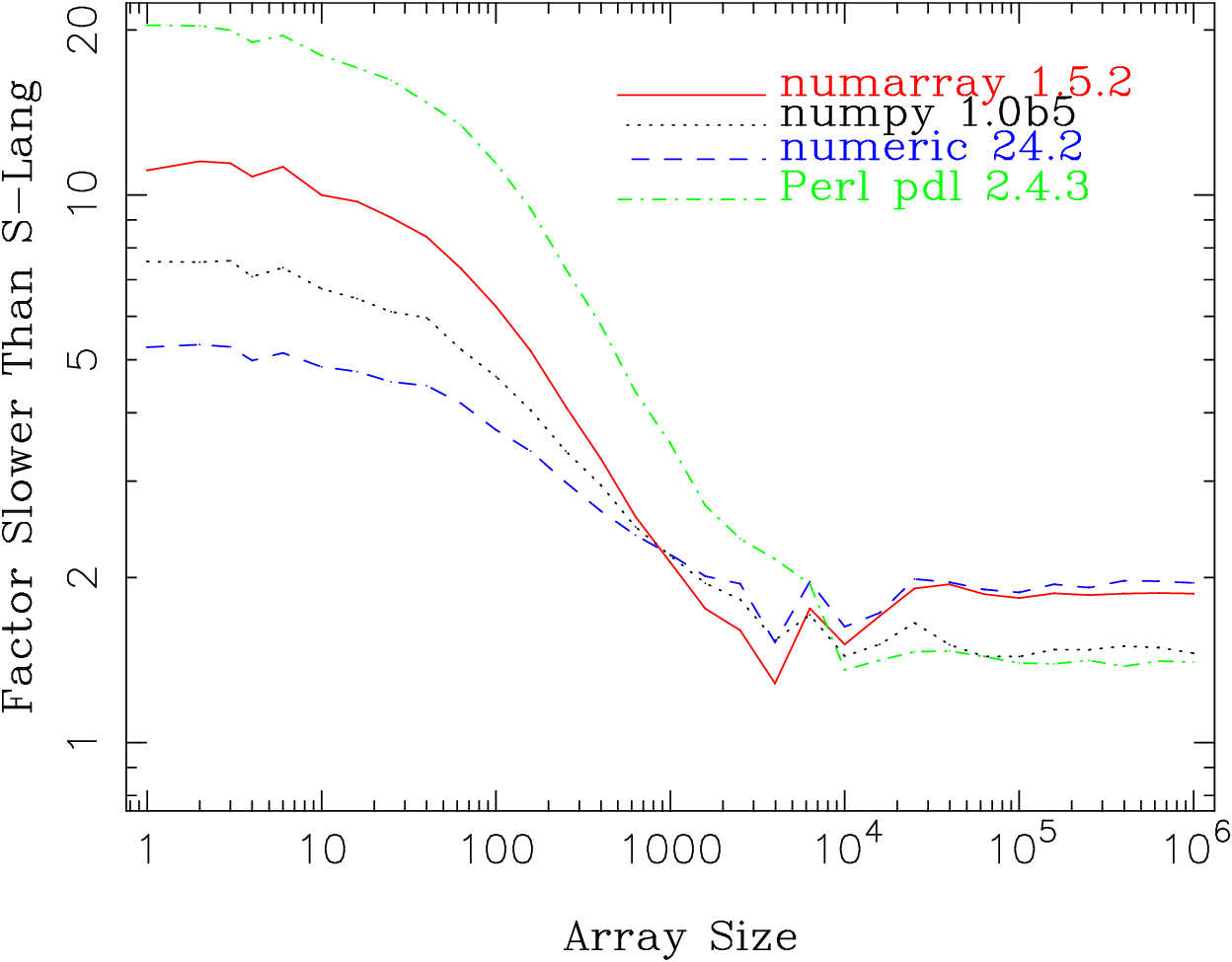}
  \hspace*{2mm}
  \includegraphics[angle=-90,scale=0.28]{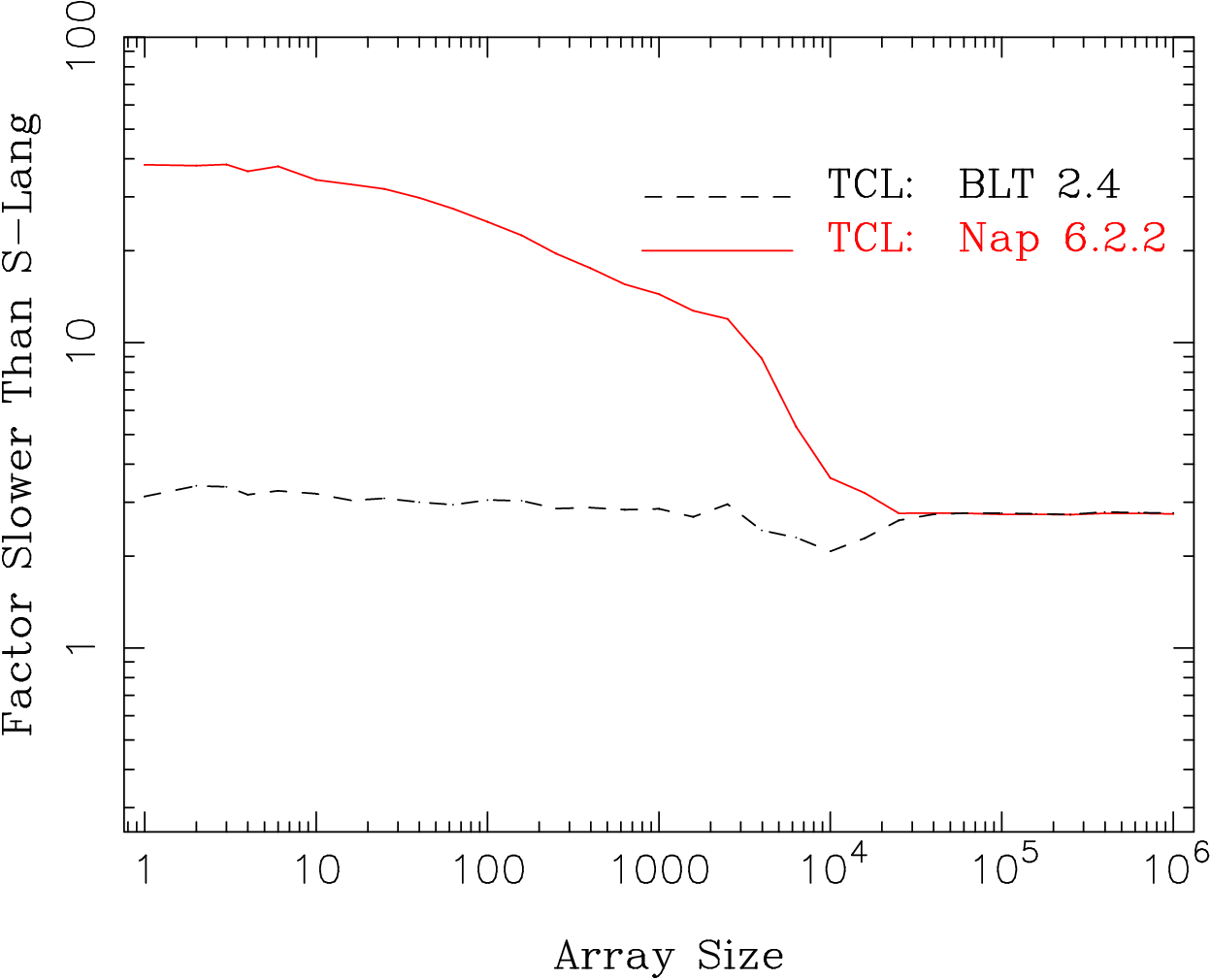}
  \caption{\lino performance of Perl, Python, and Tcl numerical
	extensions, relative to \slang, on $\sqrt{b^2 - 4ac}$, where
	a, b, and c are arrays; smaller numbers are better.}
  \label{sqrt-plot}
  \end{centering}
  \vspace*{3mm}
\end{figure*}
\begin{figure*}
  \begin{centering}
  \includegraphics[angle=-90,scale=0.28]{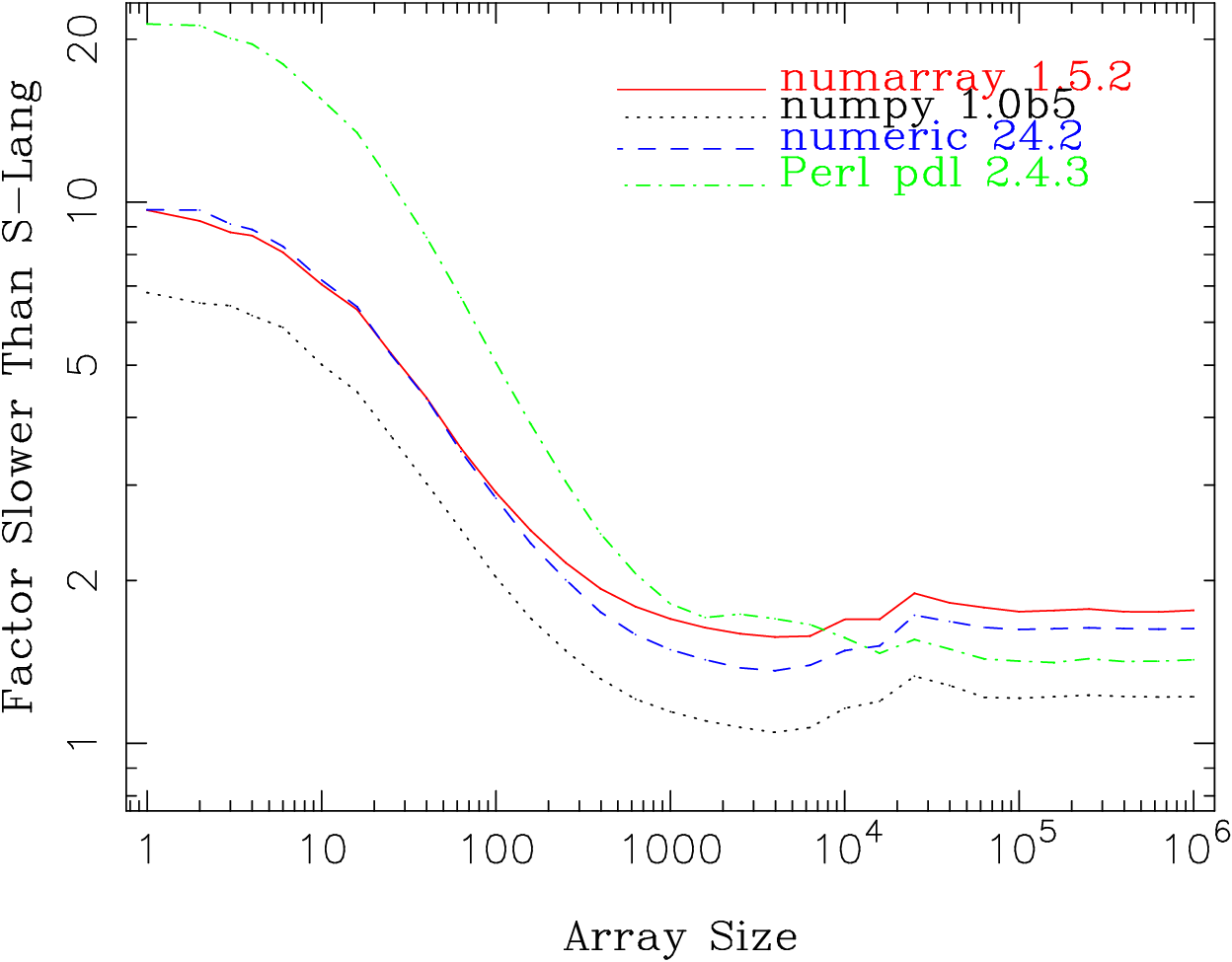}
  \hspace*{2mm}
  \includegraphics[angle=-90,scale=0.28]{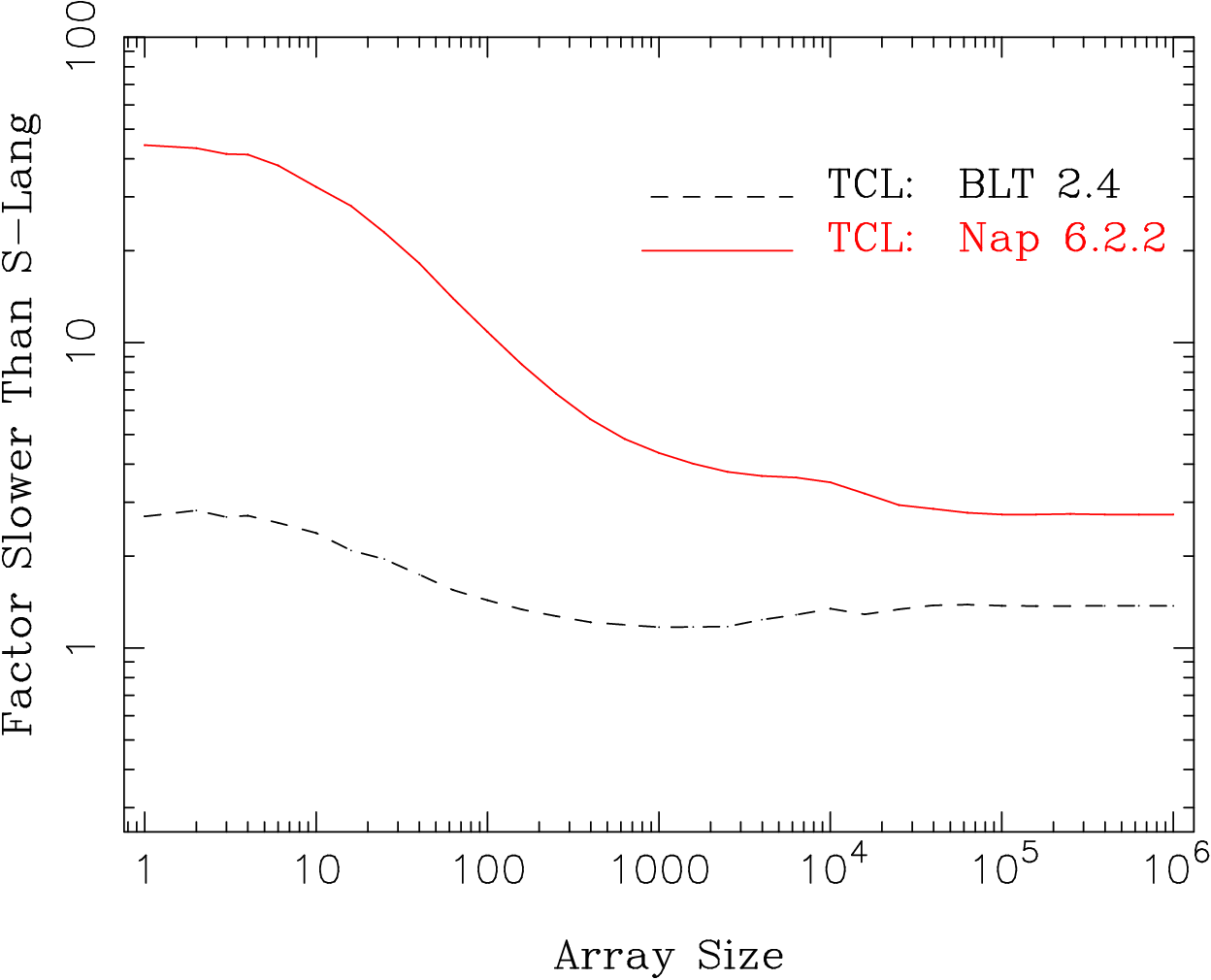}
  \caption{\lino relative performance on the array slicing expression
           $(a^{1.5} / 2 + b^2 / 4 + n^2*sin(c^3) / 5) <= n^2$,
           where n is the array size; smaller numbers are better.}
  \label{arrayexpr-plot}
  \end{centering}
\end{figure*}
\begin{figure*}
  \begin{centering}
  \includegraphics[angle=-90,scale=0.278]{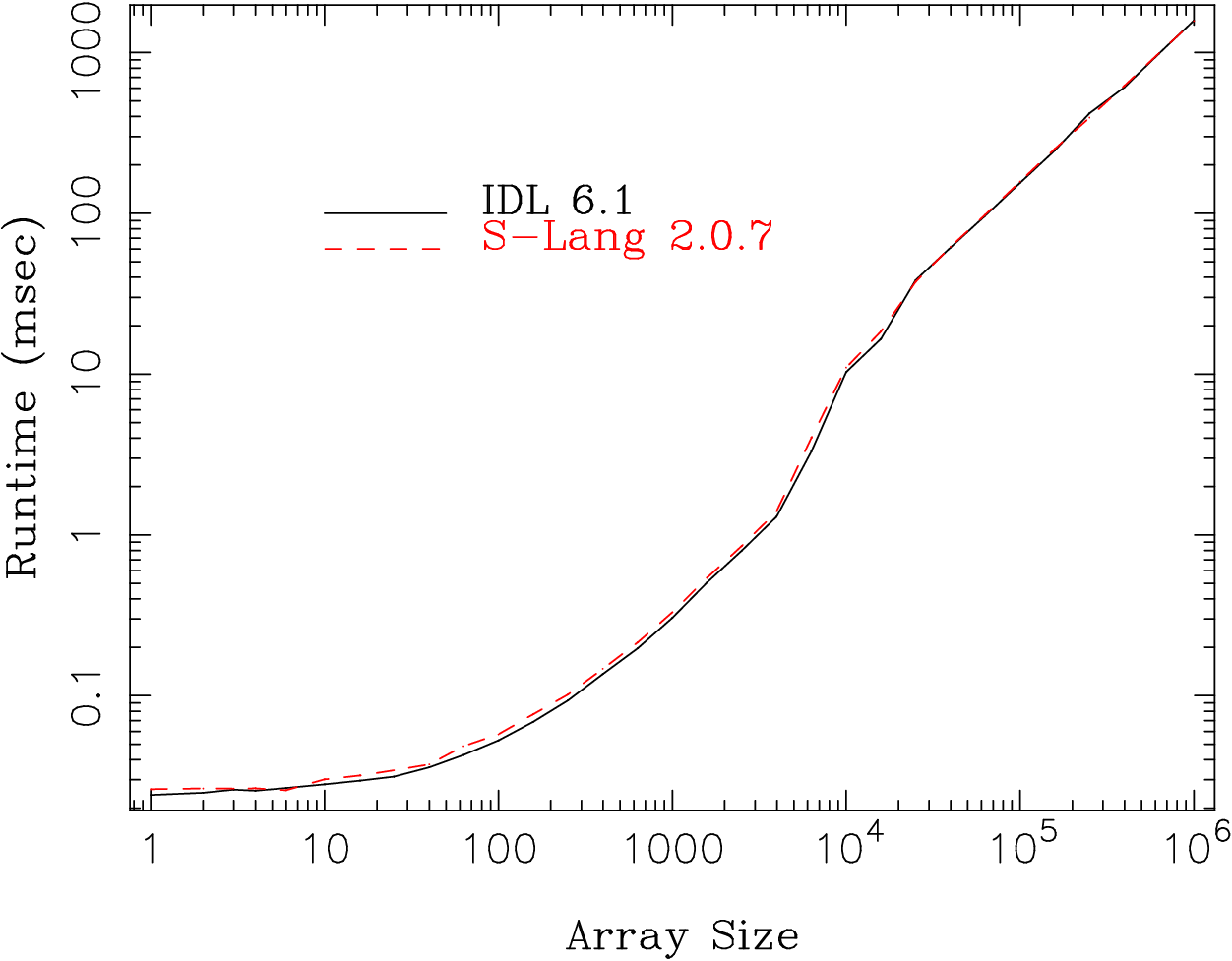}
  \hspace*{0.12in}
  \includegraphics[angle=-90,scale=0.24]{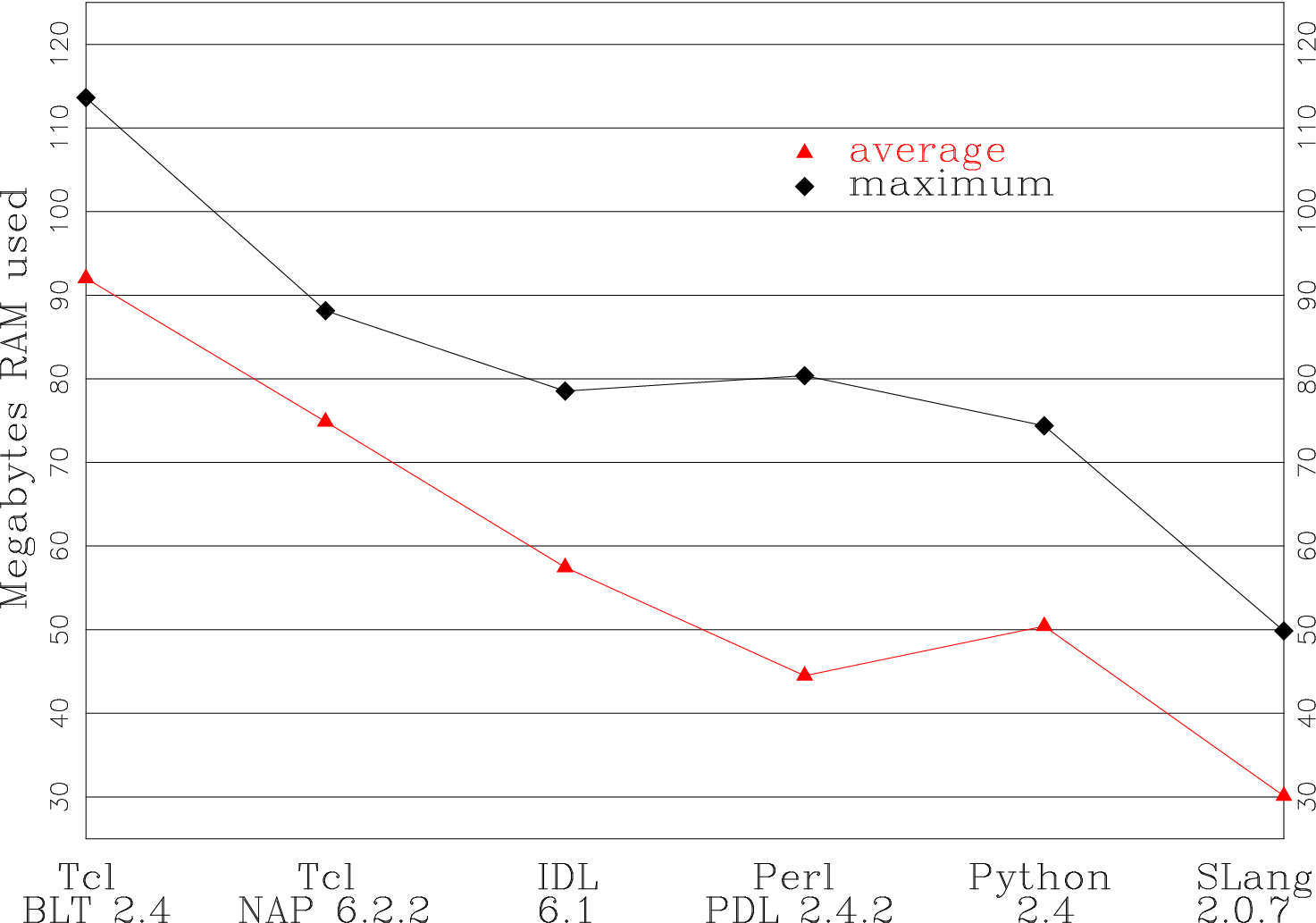}
  \caption{Left: \lino performance of IDL 6.1 (binary) and \slang
   (statically linked), for $\sqrt{b^2 - 4ac}$.
   Right: memory consumed within each language environment when
   computing $\sqrt{b^2 - 4ac}$; \hspace*{0.05in} smaller numbers
   are better.} \label{mem-plot}
  \end{centering}
  \vspace*{3mm}
\end{figure*}
numerical engine of \slang
is beyond the scope of this paper, Figs. \ref{sqrt-plot} - \ref{mem-plot}
show 2 representative calculations on Linux1 which give a rough indication
of its performance -- in terms of speed and memory utilization -- relative
to IDL and the Tcl, Perl, and Python numerical extensions.  With the
exception of IDL, which we used from a (presumably optimized) binary
installation,
all codes were compiled with GCC 3.3.5 using -O3 and -fPIC flags.  A
total of 31 datapoints were gathered per language per computation, each
representing the mean time of 1000 invocations of the respective calculation
with a given array size (from 1 to 1e6 elements),
using 8-byte real values.  Memory statistics were collected by the
{\tt proc-time} utility,\footnote{http://www.xs4all.nl/\%7Ejantien/software}
with small modifications.

\section{\slirp}
\slirp is a wrapper generator which simplifies the creation of dynamically
loadable modules  for \slang.  It can dramatically reduce the
effort needed to make external C/C++ and Fortran codes callable directly
from the \slang interpreter.  \slirp is implemented in \slang and a small
amount of supporting C code, with no other external dependencies, making it
easy to install, use, and rapidly evolve.  It has been used to generate
bindings to a wide range of software, from libraries as large as Gtk and
HDF5 to codes
as small as numerical models for \isis contained within a single file.
It is particularly useful as a means of quickly assessing whether a piece
of software is a good candidate module for \isis; our group has evaluated
numerous codes in this manner.

\subsection{Vectorization}
Perhaps the most distinguishing feature of \slirp is its ability to
vectorize wrapped functions, optionally tuned for parallelization with
\omp.
A stark example of the power of vectorization is given in Fig. \ref{atof_perf},
\begin{figure}[t]
\begin{minipage}{0.9\textwidth}
\begin{Verbatim}[xleftmargin=1cm]
isis> avol = array_map(String_Type, &sprintf, "%d", [1:100*100*100])
isis> tic; dvol = array_map(Double_Type, &atof, avol); toc
13.754

isis> import("atof")
isis> tic; pdvol = atof(avol); toc
0.1442
\end{Verbatim}
\end{minipage}
\caption{\lino snapshot of \slang \atof versus \slirp vector-parallel
version, on $100^3$ strings.}
\label{atof_perf}
\end{figure}
which was motivated by the desire to visualize a 320x320x320
cube\footnote{See {\tt volview} at http://space.mit.edu/hydra/implement.html}
representing Doppler velocity mappings of Silicon II infrared emission
observed with the Spitzer telescope.  The 130Mb volume was supplied in
ASCII form, so to minimize I/O time during exploratory analysis we first
converted it to the
high-performance HDF5 binary format, which involved some $320^3$ calls
to \atof.  This function is not vectorized in \slang, so to boost
performance we generated a vector-parallel replacement, in only seconds
with \verb+slirp -make -openmp atof.h && make+, using \slirp to also
generate the Makefile.  With faked data scaled down for didactic expedience
to contain only $100^3$ voxels, Fig. \ref{atof_perf} indicates that the
vector-parallel version is ca. 95X faster on our \lint machine.
It is worth noting that {\tt array\_map()}
is normally the fastest technique by which non-vectorized \slang
intrinsics can be applied to arrays.  Using other iterative mechanisms,
say a {\tt for} loop, would have yielded an even greater advantage for
the parallelized \atof.
This snapshot also hints at another significant benefit of vectorization,
namely brevity of end-user code.  As a stronger example, consider the problem
of reading N datasets from an HDF5 file \cite{1999IEEE...99...273F}, where
each dataset is a 100-element single precision floating point array whose
name is stored at index {\em i} of a string array.  The minimal user-level
code to read these data into a 2D array in IDL 6.1, without resource leaks,
is given in the left panel of Fig. \ref{hdf5};  the same result can be
achieved with our vectorized HDF5 module in a single statement.
\begin{figure}[h]
 \begin{minipage}{62mm}
 \begin{Verbatim}
 array = fltarr(N, 100)
 fp = H5F_OPEN(file)
    for i = 0, N-1 do begin
        dp = H5D_OPEN(f, datasets[i])
           array[i, *] = H5D_READ(dp)
        H5D_CLOSE(dp)
    endfor
 H5F_CLOSE(fp)
 \end{Verbatim}
 \end{minipage}
 \hspace{2mm}
 \begin{minipage}{62mm}
  \begin{Verbatim}




   array = h5_read(file, datasets);



  \end{Verbatim}
 \end{minipage}
 \caption{Reading multiple HDF5 datasets with IDL 6.1 (left) and the vectorized
 \slang module.}
 \label{hdf5}
\end{figure}

Vectorization encompasses
more than the simple promotion of scalar arguments to arrays. More
generally, we say a function is vectorized when its arguments may
be of multiple ranks. No distinction is made between the promotion of a
rank 0 scalar to 1D or higher, a 2D array to 3D, and so forth.  When
a vectorized function is invoked with any argument whose rank exceeds
that of its prescribed usage we
say that both the argument and the function call are {\em vectored}.
\slirp aims for maximum flexibility, allowing vectorized
functions to be invoked using either scalar or array semantics and with few
restrictions on the quantity, datatype, or dimensionality of arguments.
For example, Fig. \ref{vmult} shows a call mixing non-isomorphic arrays.
The wrapped C function is prototyped as
\begin{verbatim}
       void vmult(double *x, double *y, double *result, int len);
\end{verbatim}
to multiply 2 vectors of length {\tt len} and is called from \slang as
\begin{verbatim}
                double[] = vmult(double[], double[])
\end{verbatim}
The two signatures differ because an {\em annotation}\footnote{Annotations
are similar to SWIG {\em typemaps}, and are described in the \slirp
documentation.} has been applied to make the \slang usage more natural:
{\tt result} is moved from the parameter list to become a return value, and
the vector length parameter is omitted because it can be obtained by
inspecting the \slang arrays.  The first \vmult call fails for the obvious
reason that vectors of dissimilar
length cannot be multiplied.  The second call succeeds, but is not vectored
because the ranks of both arguments match those of their prototyped parameters.
The final call is vectored because the rank of the first argument, a 2D
array, exceeds its prototyped dimensionality of 1.
\begin{figure}[t]
 \begin{minipage}{62mm}
  \begin{Verbatim}
isis> vmult([1,2,3], [3,4])
Array shape or length mismatch
isis> print( vmult([1,2,3], [5,5,5]) )
5
10
15
  \end{Verbatim}
 \end{minipage}
 \hspace{.25mm}
 \begin{minipage}{62mm}
  \begin{Verbatim}
isis>  Arr = Double_Type[2,3]
isis>  Arr[0,*] = 5
isis>  Arr[1,*] = 100
isis>  print( vmult(Arr, [3, 4, 5]) )
15 20 25
300 400 500
  \end{Verbatim}
 \end{minipage}
 \caption{Invoking a vectorized function with arrays of both similar 
 and dissimilar shapes.}
 \label{vmult}
\end{figure}
\subsection{Dimensionality Theory}
\slirp uses a few simple metrics to decide whether a wrapper has been
called with vectored semantics, collectively referred to as the
{\em parameters of vectorization}.
To begin, each argument passed to a wrapper has an {\em expected rank}: a
non-negative integer indicating the number of indices required to
uniquely identify a single element.  This rank is inferred at code
generation time from the arguments signature within the function
declaration.  \slirp 
distinguishes dimensioned arrays such as \verb+double x[3][5]+ from
arrays of pointers like \verb+double **x+, assigning them ranks of 2
and 1, respectively.

The {\em actual rank} of an argument is its dimensionality as passed
at runtime.  When the actual rank of any argument exceeds its expected
rank, \slirp needs to determine how many times the wrapped function should
be called, or the {\em number of iterations} of the vectorization.  This
is decided by selecting a master array \M \xspace-- the input argument of
highest rank -- and computing the product of its excess dimensions.  For
example, if {\tt Arr} in Fig. \ref{vmult} was 4x3x3 instead of 2x3 then
\vmult proper would be called 12 times instead of 2.  Formally,
if \A and \E represent the actual and expected ranks of \M, and \D is a
vector of length \A describing the size of each dimension of \M (in
row-major form), then
\begin{equation}
  Num\_Iterations = \left\{
	\begin{array}{ll}
	   1	& \mbox{when \A = \E} \\
	   \displaystyle{\prod_{i=1}^{A-E}{D[i]}} & \mbox{when \A $>$ \E}.
	\end{array}\right.
\label{numiters}
\end{equation}
Finally, \slirp determines what to pass to the wrapped function by
calculating a {\em stride} for each argument; this indicates by how
much an index into the argument -- viewed as a linear sequence of
contiguous elements -- should be advanced after iteration of the
vectorization loop.  Returning to the \vmult call in Fig. \ref{vmult},
the strides of the first and second arguments are 3 and 0; within the 
wrapper the input arguments and return value are effectively represented as
\begin{Verbatim}[frame=none,xleftmargin=3cm]
double  *arg1 = {5, 5, 5, 100, 100, 100};
double  *arg2 = {3, 4, 5};
double  *retval = malloc( sizeof(double) * 6);
\end{Verbatim}
and the 2 calls to \vmult proper are executed as
\begin{Verbatim}[frame=none,xleftmargin=3cm]
vmult(arg1, arg2, retval, 3);
vmult(arg1+3, arg2+0, retval+3, 3);
\end{Verbatim}
Formally, the stride of \M and all isomorphic arguments is the number of
elements contained within its expected dimensions
\begin{equation}
  Stride \mspace = \mspace \prod_{i=A-E+1}^{A}{D[i]}.
\end{equation}
The stride can be computed directly from the number of iterations by
recalling that the number of elements in \M is the product of its dimensions
\begin{equation}
  Num\_Elements \mspace = \mspace \prod_{i=1}^{A}{D[i]}.
\end{equation}
\noindent
Factoring the left side into the product of excess and expected dimensions gives
\begin{equation}
  Num\_Elements = \mspace \prod_{i=1}^{A-E}{D[i]} \prod_{i=A-E+1}^{A}{D[i]},
\end{equation}
and by noting that the first term here is the number of iterations we see
\begin{equation}
  Stride \mspace = \mspace Num\_Elements \mspace  / \mspace Num\_Iterations.
\label{stride2}
\end{equation}
Equations \ref{numiters} and \ref{stride2} are coded into the {\tt vec\_pop()}
routine discussed in the next section.
Arguments not isomorphic to \M are legal as long as their number of
elements equals the stride of \M; they will be assigned a stride of 0.

\subsection{Anatomy of a Vectorized Wrapper}
To give a sense of what vector-parallelism entails, Fig.
\ref{hypot_wrappers} shows the code generated for vectorized and parallel
wrappers of the C \Hypot function.
In the vectorized wrapper the return value and arguments of \Hypot
are pointers, instead of scalars as they would be in a standard wrapper,
with additional reference variables declared
to record argument metadata such as array dimensions and stride.  Two
additional variables are declared to support vectorization: a scalar to
index the vectorization loop, and a {\tt VecSpec} structure to reflect the
parameters of vectorization, which are adjusted by {\tt vec\_pop} as it
marshals arguments from \slang.
\begin{figure}
(a)
\begin{minipage}{.98\textwidth}
\begin{Verbatim}[xleftmargin=2mm]
static void sl_hypot (void)
{
   double* retval;
   double* arg1;
   Slirp_Ref *arg1_r = ref_new(SLANG_DOUBLE_TYPE,sizeof(double),&arg1,0x0);
   double* arg2;
   Slirp_Ref *arg2_r = ref_new(SLANG_DOUBLE_TYPE,sizeof(double),&arg2,0x0);
   unsigned int _viter;
   VecSpec vs = {1, 0, 0};

   if (SLang_Num_Function_Args != 2 ||
	vec_pop( arg2_r, 0, 0, &vs) == -1 ||
	vec_pop( arg1_r, 0, 0, &vs) == -1 )
	{ Slirp_usage(0,0,1); finalize_refs(VREF_2); return; }

   if (vec_validate(&vs, VREF_2) == -1) {finalize_refs(VREF_2); return;}
   VEC_ALLOC_RETVAL(double, VREF_2);
   for (_viter=0; _viter < vs.num_iters; _viter++) {
	retval[_viter] = hypot(*arg1,*arg2);
	VINCR_2;
   }
   VEC_RETURN(retval, 0, SLANG_DOUBLE_TYPE, SLang_push_double, 0, 1);
   finalize_refs(VREF_2); 
}
\end{Verbatim}
\vspace*{2mm}
\end{minipage}
(b)
\begin{minipage}{.98\textwidth}
\begin{Verbatim}[xleftmargin=2mm]
static void sl_hypot (void)
{
   ...
   int _viter;
   VecSpec vs = {1, 2, 0};

   ...
	{ Slirp_usage(0,0,3); finalize_refs(VREF_2); return; }
   ...

   #pragma omp parallel for
   for (_viter=0; _viter < vs.num_iters; _viter++) {
	retval[_viter] = hypot(arg1[_viter],arg2[_viter]);
   }
   ...
}
\end{Verbatim}
\end{minipage}
\caption{Vectorized (a), and parallelized (b) wrappers for \Hypot.
}
\label{hypot_wrappers}
\end{figure}
The {\tt vec\_validate} function ensures that \Hypot can safely be called
with the given inputs; it may also adjust the stride of non-isomorphic
arguments along the way and allocate space for arguments which have
been omitted from the \slang wrapper invocation, such as {\tt double *result}
from \vmult, because they are still required
by the wrapped function.  The {\tt VEC\_ALLOC\_RETVAL}, {\tt VEC\_RETURN},
{\tt VREF\_}{\em n}, and {\tt VINCR\_}{\em n} macros enhance readability
by masking unnecessary detail, performing tasks like memory management,
argument striding, and the expansion of argument lists for support routines.

Serial vectorization appeared first in \slirp, but Fig.
\ref{hypot_wrappers}-(b) shows how easy it was to adapt for \omp: most of
the parallel wrapper is identical to the serial version and has been elided.
In addition to the \omp pragma -- and the usage message which was changed
to indicate that the wrapper is also parallelized, the loop index has been
changed to a signed integer. This unfortunately reduces by half the maximum
size of array operations which may be parallelized, and also led to more
extensive code
changes during the operator parallelizations described in
\S \ref{parallel-opers}, but was necessary for conformance with the \omp
2.5 specification; we look forward to the support for unsigned indices
coming in \omp 3.0,
Note that array indexing is
used to locate elements, instead of pointer dereferencing.  Pointers are
used in serial wrappers for flexibility -- they cleanly enable each argument
to have its own stride. In \omp loops, however, pointer traversals introduce
unwanted concerns for portability and possible non-conformance with
the specification.  Although techniques exist for iterating over pointers
within \omp blocks \cite{1131942}, we avoid the additional complexity and
potential performance degradation by using a single loop variable to index
all arguments; the tradeoff is that all arguments must be isomorphic, enabling
the same stride to be used for each.

\begin{figure}[t]
\begin{Verbatim}
define weibull_fit(lo, hi, params)
{
   variable a, b, x1, r=@lo, i, c, m, d, e;

   a = params[0]; b = params[1]; c = params[2]; d = params[3];
  
   x1 = c - b * ((a-1)/a)^(1/a);
   m = (hi+lo)/2.0;

   i = where(lo > x1);
   if(any(i))
	r[i] = d*(a/b)*((m[i]-x1)/b)^(a-1)*exp(-((m[i]-x1)/b)^a);

   i = where(lo <= x1);
   if(any(i))
	r[i] = 0;

   return r;
}
\end{Verbatim}
\caption{The 4-parameter Weibull model in \slang, as a custom fit function for \isis.  The {\tt lo} and {\tt hi} arrays represent bin edges in a 1D grid; their
sizes vary as the X axis value in Fig. \ref{fit-perf}.}
\label{weibull}
\end{figure}

\section{Performance Experiments}
\label{performance}

In addition to \atof and \Hypot wrappers were generated for the
\Sin, \Cos, \Exp and \Log intrinsics.  We chose to evaluate functions
already available and vectorized in \slang to illustrate several
points:  First, that \slirp introduces minimal overhead; second, that
\slang intrinsics can be replaced at runtime, without altering
\noindent
the semantics of numerical scripts\footnote{With one exception: complex
types are not supported
by the wrappers parallelized here. \slirp {\em can} wrap Fortran codes
with complex arguments and return values, however.};  and finally, that
this leads to the use of existing serial algorithms in parallel contexts,
without recoding.  Consider for
example the \slang function in Fig. \ref{weibull}, which defines a Weibull
model \cite{Weibull.1939}
for fitting in \isis.  While Fortran and C/C++ models may also be imported
into \isis, in just seconds with \slirp, it can be
faster to code them directly in \slang and avoid compilation steps
during experimental tuning. The high performance numerics in \slang 
means that such interpreted models need not trade speed for convenience.
\begin{figure}[t]
  \rotatebox{-90}{\scalebox{0.27}{\includegraphics{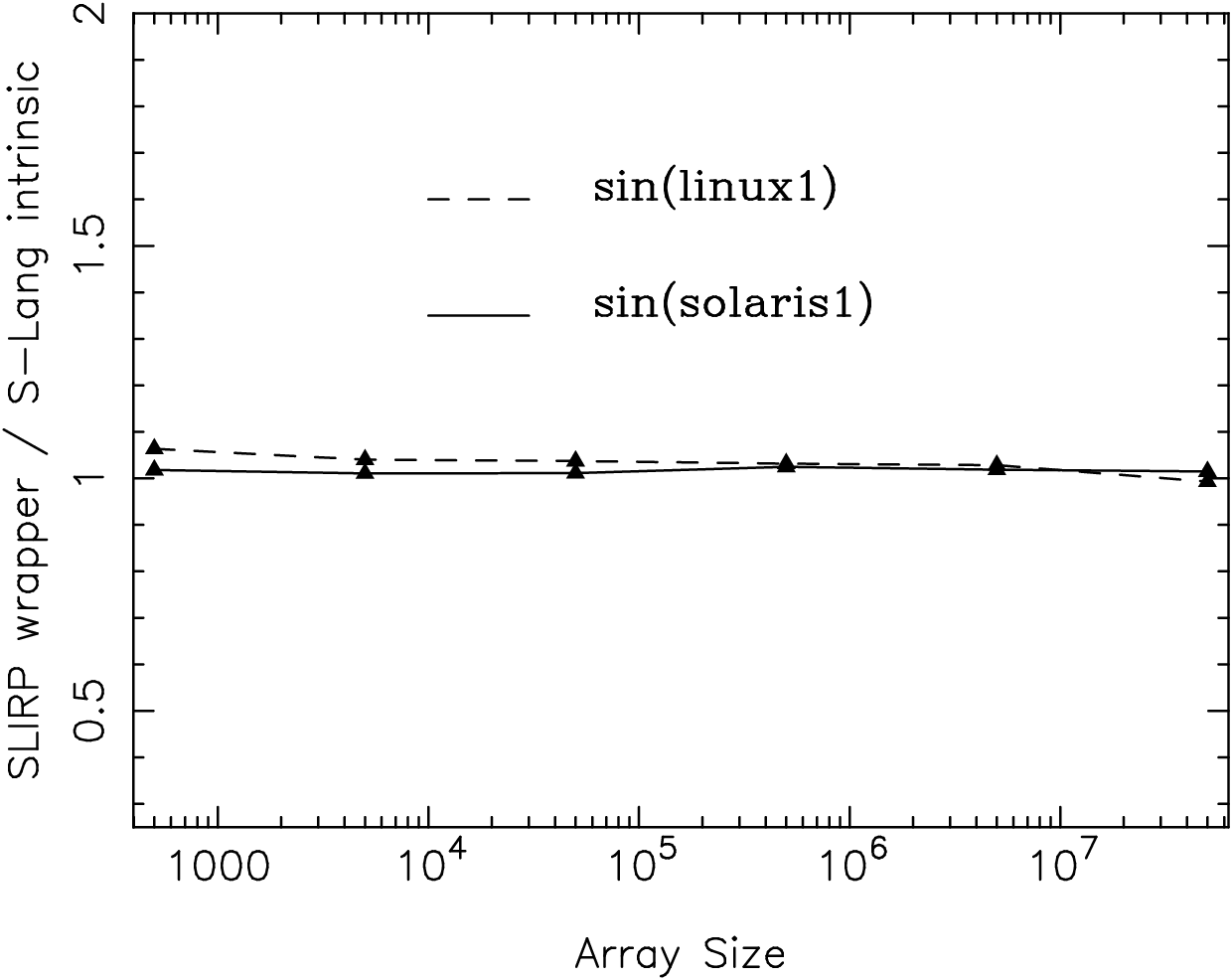}}}
  \hspace*{10mm}
  \rotatebox{-90}{\scalebox{0.27}{\includegraphics{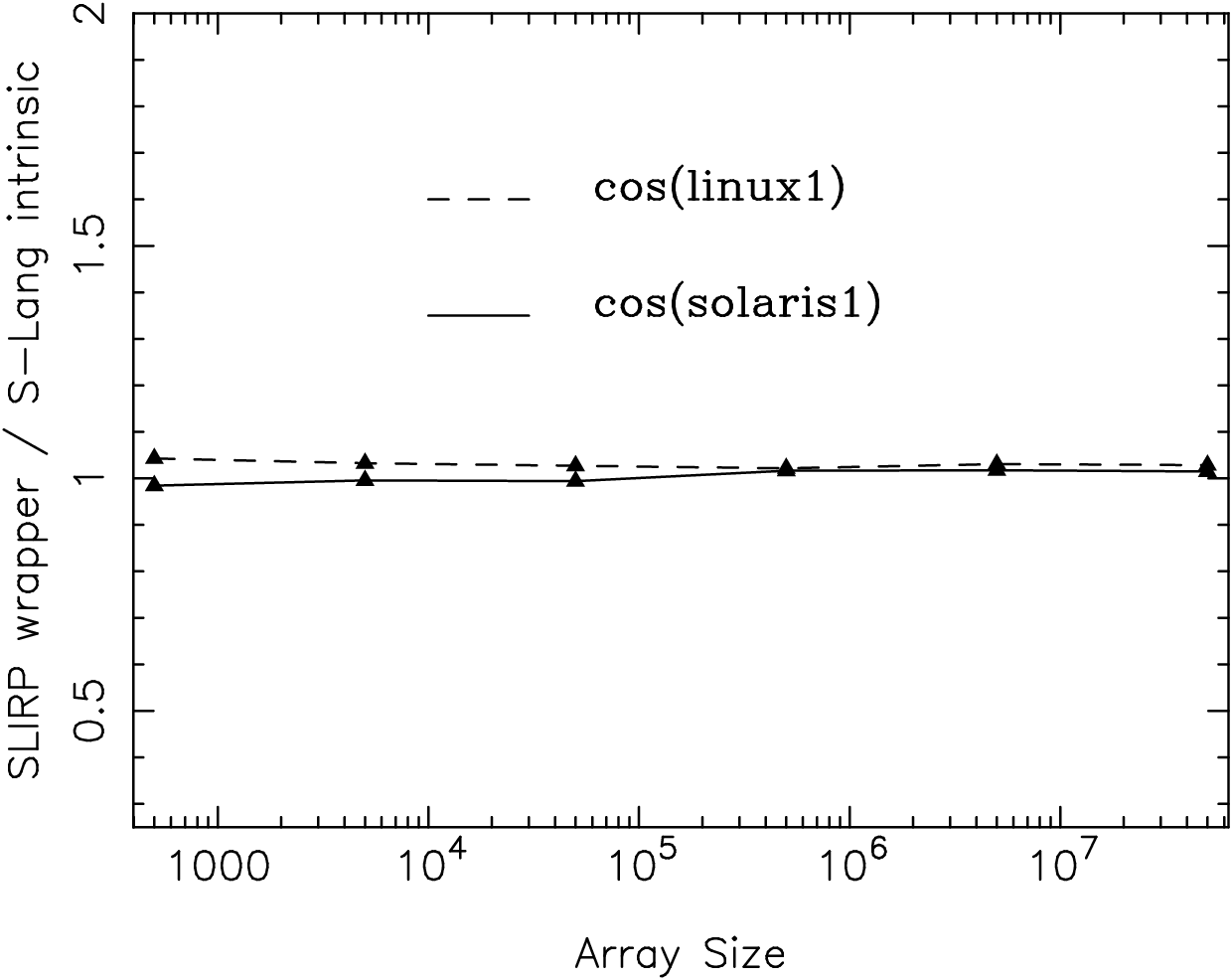}}}
 \caption{Serial performance ratios of \slirp-vectorized wrappers of
 {\tt sin} and {\tt cos} versus hand-crafted \slang intrinsics, on \lino
 and \solo.  Mean runtimes per array size ranged from ca. 3.2e-5
 to 4.2 seconds on \lino, and from 2.2e-4 to 9.15 seconds on \solo.}
 \label{serial-plots}
\end{figure}
\subsection{Parallel Functions}
This model was taken from an active research project and originally
coded for serial use; it was parallelized in two ways, neither of
which involved changing a single line of its code.
First, by using \verb+import("par")+ to dynamically load the module
of parallel wrappers generated by \slirp.  This can be done either
interactively at
the command prompt (as in Fig. \ref{atof_perf}), programmatically in
a script, or even automatically at \isis launch by putting the
{\tt import()} within an {\tt .isisrc} initialization file.  The
advantages of using \slirp for parallelization are automation and
simplicity: it may be employed immediately in \slang
2 applications by any user, merely by installing an \omp-aware compiler;
having \omp support in GCC now makes this a much lower barrier than in
the past.
\subsection{Parallel Operators}
\label{parallel-opers}
A shortcoming of this tactic, however, is that it can only be used to
parallelize functions, leaving a potential vulnerability to Amdahl's
Law: in an
expression such as \verb@cos(x)/5 + sin(x)/2@ the two divisions and one
addition would still be computed serially.
Our second approach to multiprocessing therefore involved manually
parallelizing the $+$, $-$, $*$, $/$,
$<=$, $>$, and \verb+^+ (exponentiation) operators by adding
\begin{verbatim}
      #pragma omp parallel for if (size > omp_min_elements)
\end{verbatim}
to the operator loops as discussed in \S \ref{openmp};  the \where function
was partially parallelized, too, and utilized an additional {\tt reduction}
clause.
Although parallel operators increase performance, a disadvantage of this
approach is that it requires edits to the internals of \slang and these
changes are not yet available to the general public.  The {\tt if} clause
in the \omp directives was used to tune performance for small array sizes,
where the cost of threads outweighs the serial execution time. During 
measurement the control variable was set with {\tt getenv()} to one of
the values \verb+{0, 500, 1000, 5000, 10000, 50000, 100000}+.

\subsection{Results and Analysis}
\label{analysis}
Unless otherwise noted, the plots discussed here represent
measurements of prerelease GCC 4.2 -O2 builds on Linux2 and Sun Studio
9 -xO3 builds on Solaris4, with position independent compilation.
Comparable trends were seen in additional testing with
the Intel 9.1 compiler on \lint and prerelease GCC 4.3 on an
2.33 Ghz Intel Core Duo Macintosh laptop (3 GB RAM) running OS/X 10.4.9.
Runtimes were calculated by executing each function or operator 20 times per
array size, and discarding the highest and lowest (with the exception of
Weibull) before averaging.

The ratio plots in Fig. \ref{serial-plots} indicate that the overhead
of automatic vectorization in \slirp is effectively constant and
negligible: serial \slirp wrappers are extremely competitive with the
hand-crafted intrinsics in \slang, for both large and small arrays.  Even
without parallelism \slirp vectorization therefore provides a solid path
to increased performance when wrapping external codes.  Purely serial
vectorizations of \atof and {\tt strlen}, for instance, are ca. 50X
faster than {\tt array\_map()}-ing the corresponding non-vectorized
\slang intrinsics.
\begin{figure}
  \rotatebox{-90}{\scalebox{0.27}{\includegraphics{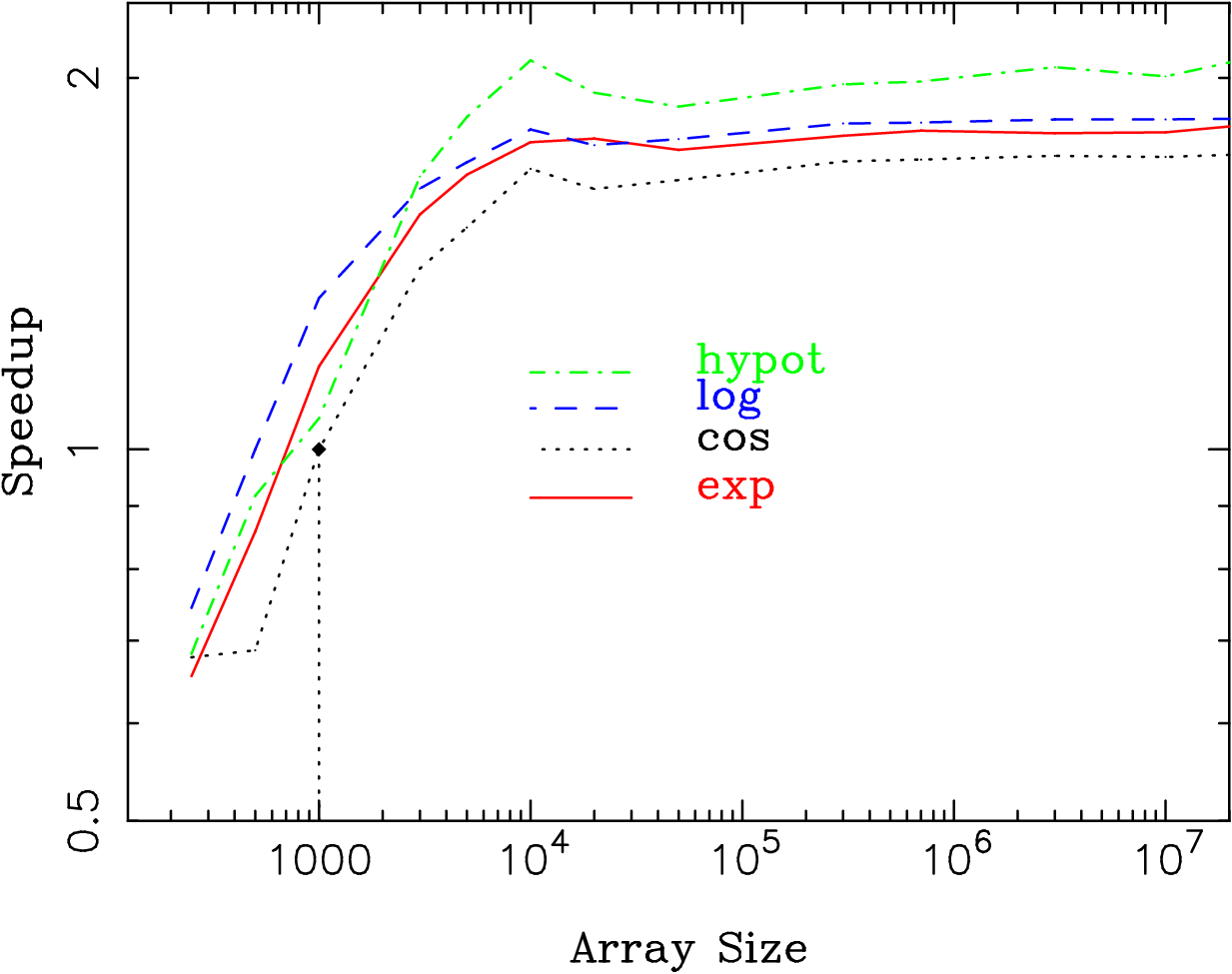}}}
  \hspace*{0.22in}
  \rotatebox{-90}{\scalebox{0.27}{\includegraphics{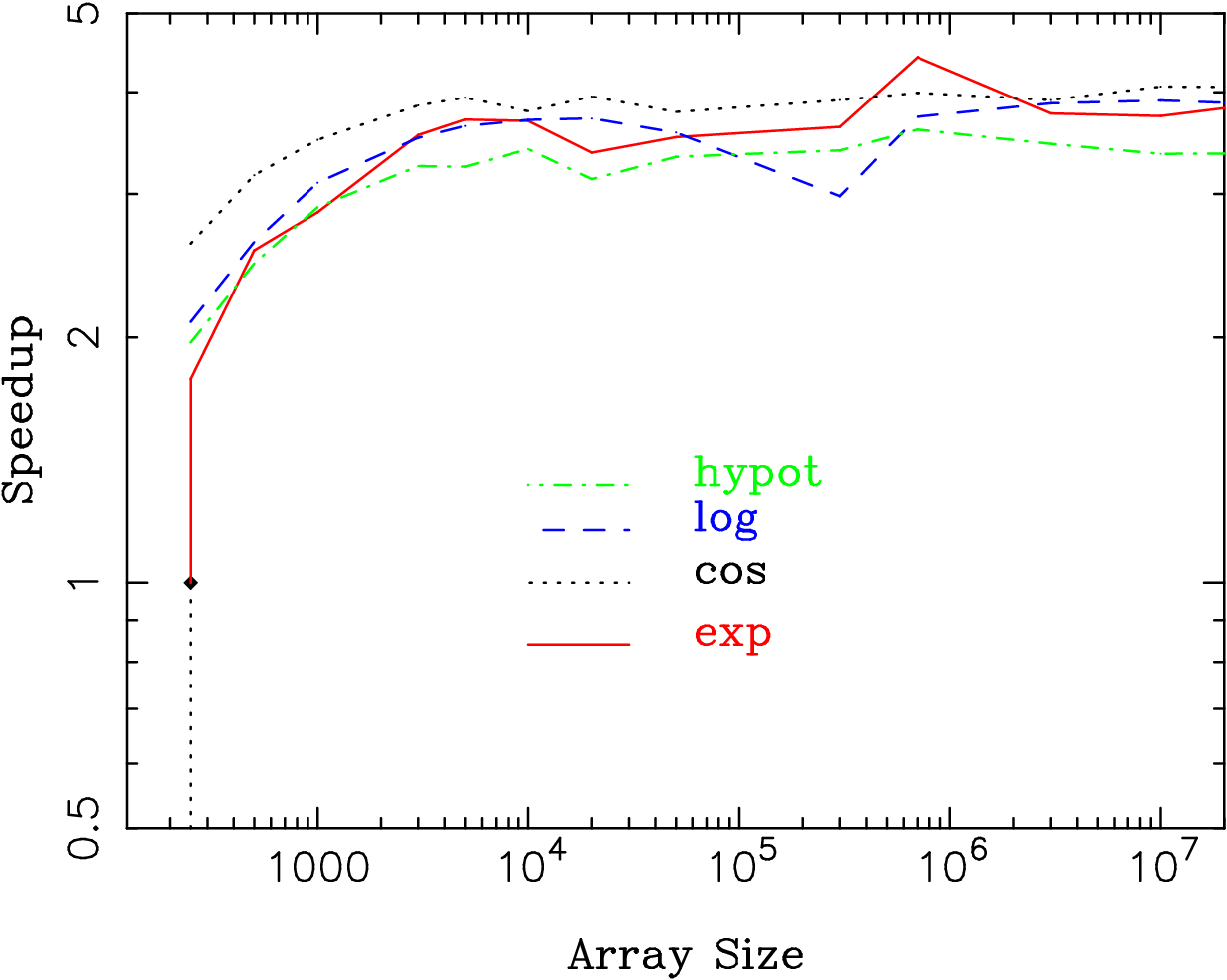}}}
  \caption{Speedups from replacing selected \slang math intrinsics
  with parallelized versions generated by \slirp \xspace -openmp.
  Left: \lint.  Right: \solf.  Mean runtimes per array size ranged
  from ca. 1.7e-05 to 2.6 seconds on \lino, and from 3.8e-05 to 9.1 seconds
  on \solo.  The dotted vertical lines mark the inflection points where 
  parallel performance begins to overtake serial, ca. 1000 elements on
  \lint and 250 elements on \solf.}
 \label{funcs-perf}
\end{figure}
\begin{figure}
 \hspace*{-7mm} (a) \hspace*{1mm}
 \begin{minipage}{.40\textwidth}
 \rotatebox{-90}{\scalebox{0.27}{\includegraphics{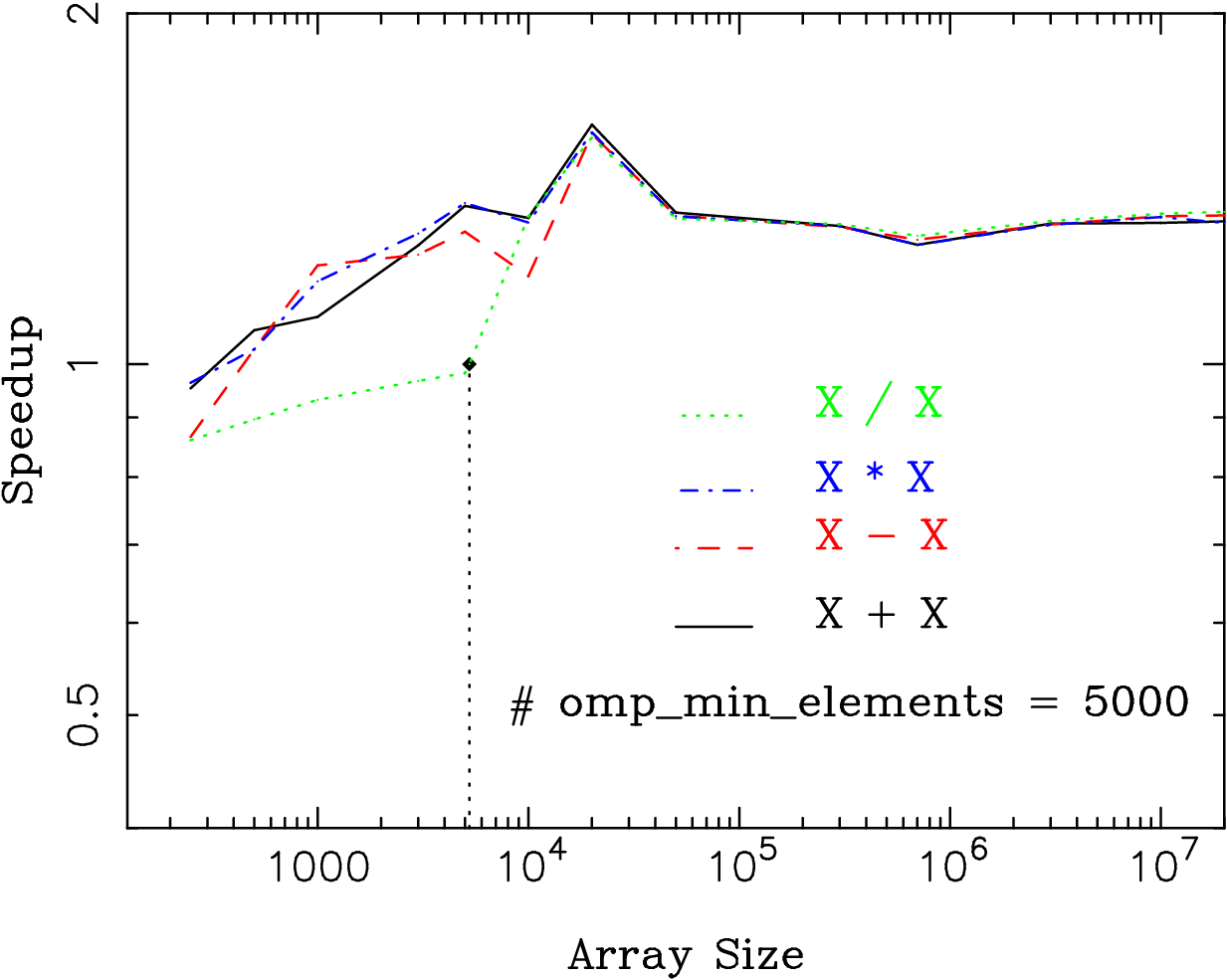}}}
 \end{minipage}
 \hspace*{11mm}
 \begin{minipage}{.40\textwidth}
 \rotatebox{-90}{\scalebox{0.27}{\includegraphics{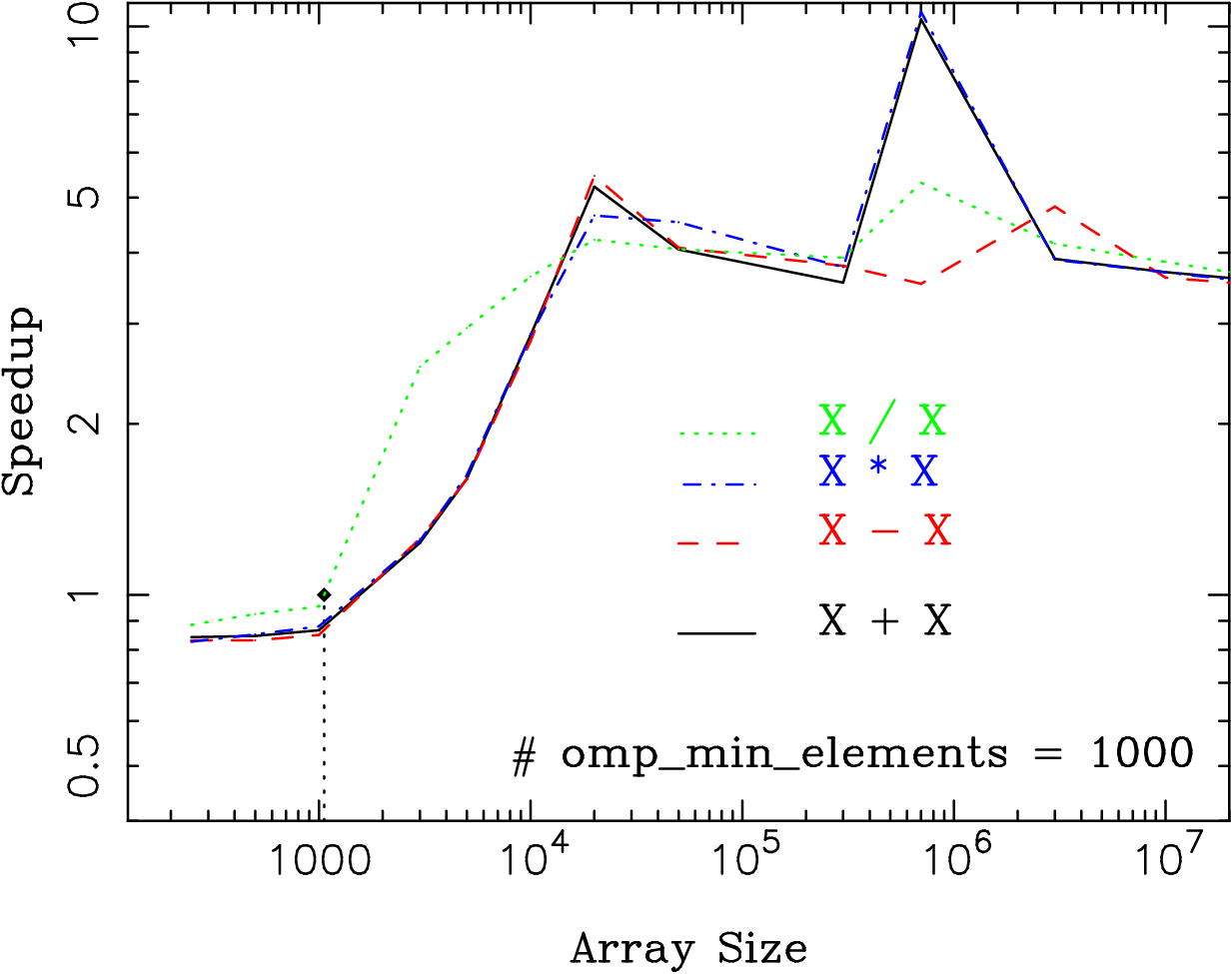}}}
 \end{minipage}
 \vspace*{7mm}
 \hspace*{6mm} (b)\\
 \hspace*{-7mm} (c) \hspace*{1mm}
 \begin{minipage}{.40\textwidth}
 \rotatebox{-90}{\scalebox{0.27}{\includegraphics{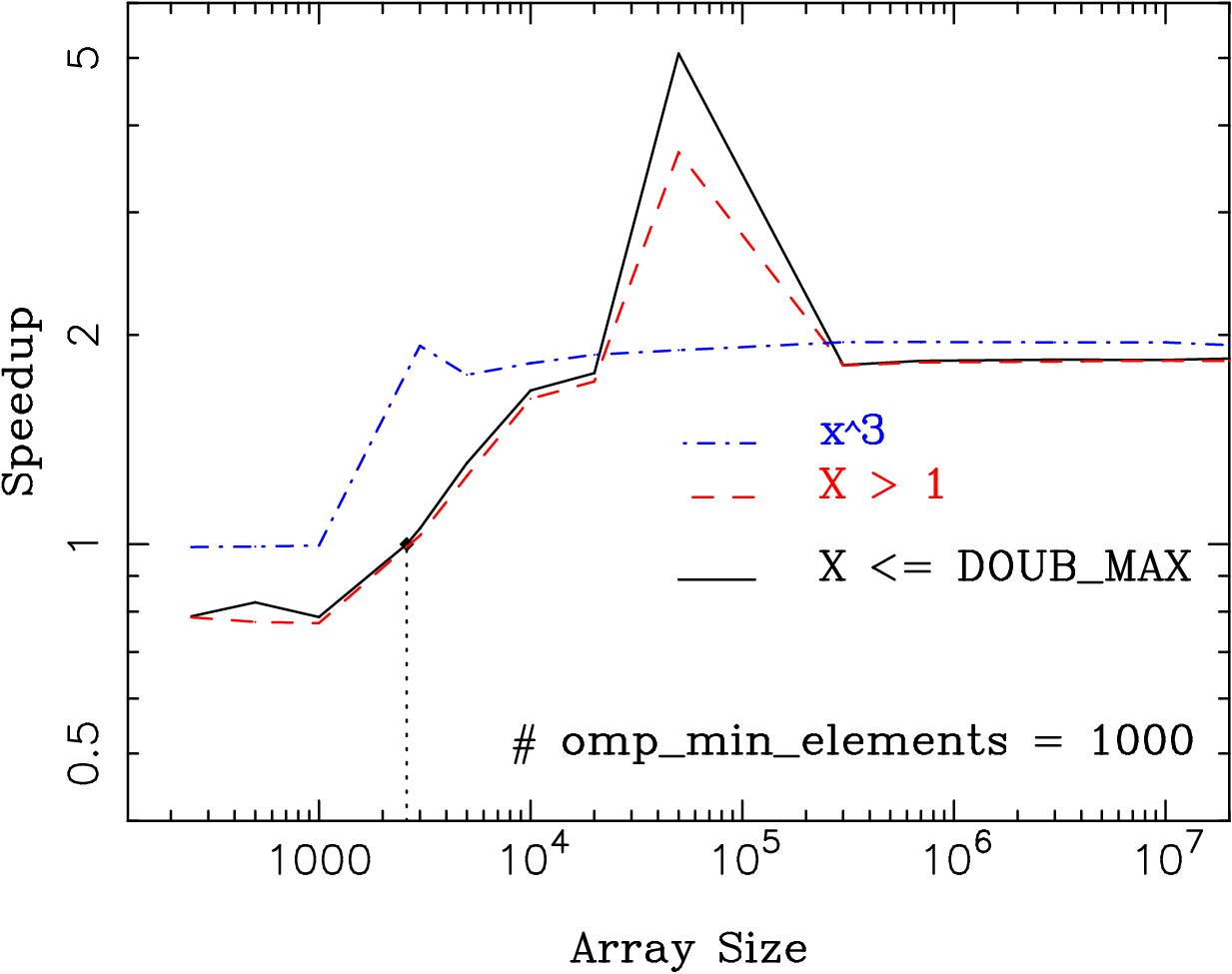}}}
  \end{minipage}
  \hspace*{11mm}
  \begin{minipage}{.40\textwidth}
  \rotatebox{-90}{\scalebox{0.27}{\includegraphics{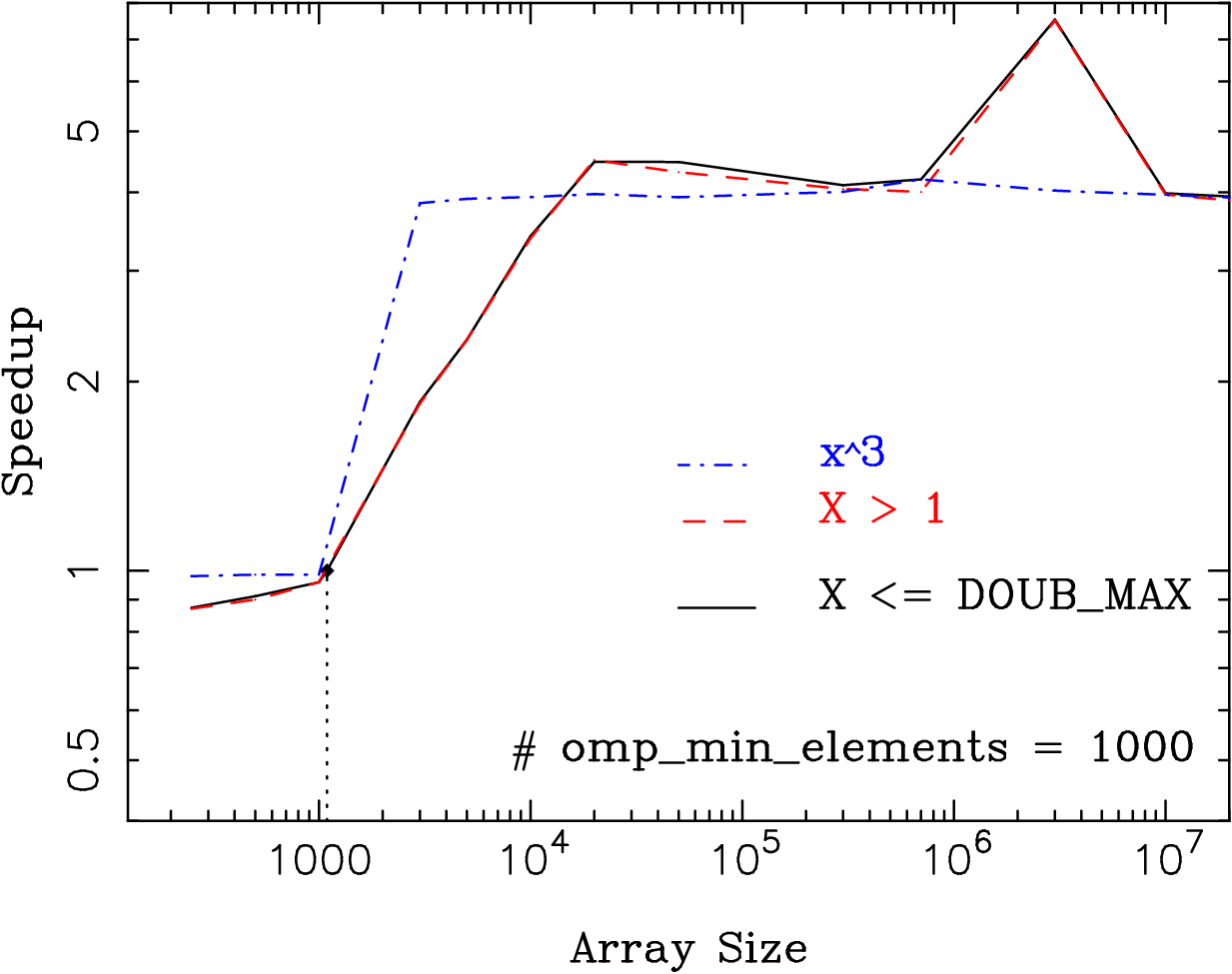}}}
  \end{minipage}
  \hspace*{6mm} (d)

  \caption{Speedups from parallelizing selected \slang operators with \omp.
  Left Top: \lint with GCC 4.2.  Left Bottom: \lint with Intel 9.1.
  Right: \solf.  \lint inflection points are at 5246 and 2602 array
  elements, with mean runtimes from ca. 3.2e-6 to 2.6 sec.
  \solf inflection points are at 1059 and 1092 elements, with
  runtimes from ca. 1.1e-5 to 13 sec.  The superlinear spikes are
  discussed in \S \ref{analysis}.}
 \label{opers-perf}
\end{figure}
\begin{figure}
  \rotatebox{-90}{\scalebox{0.27}{\includegraphics{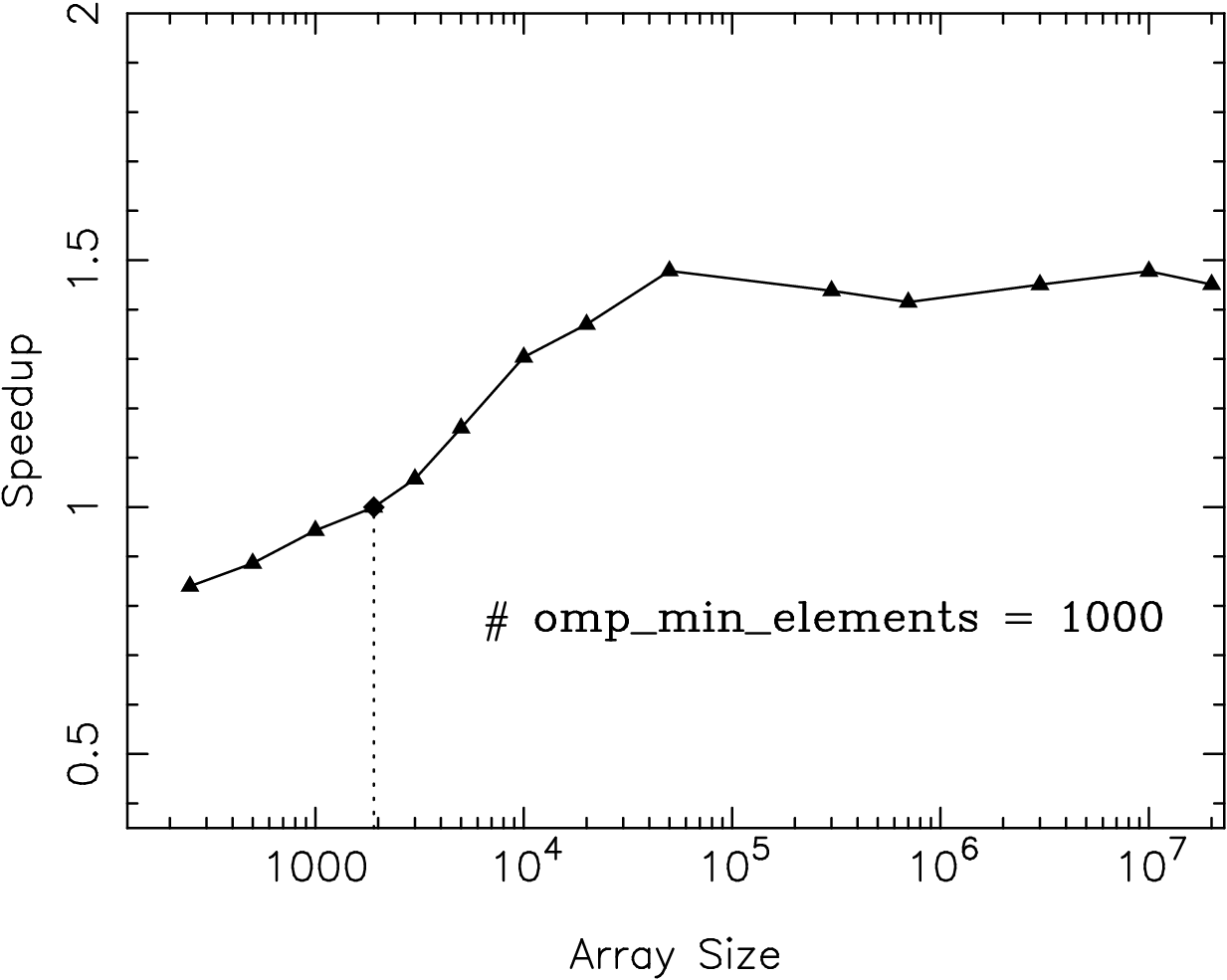}}}
  \hspace*{0.22in}
  \rotatebox{-90}{\scalebox{0.27}{\includegraphics{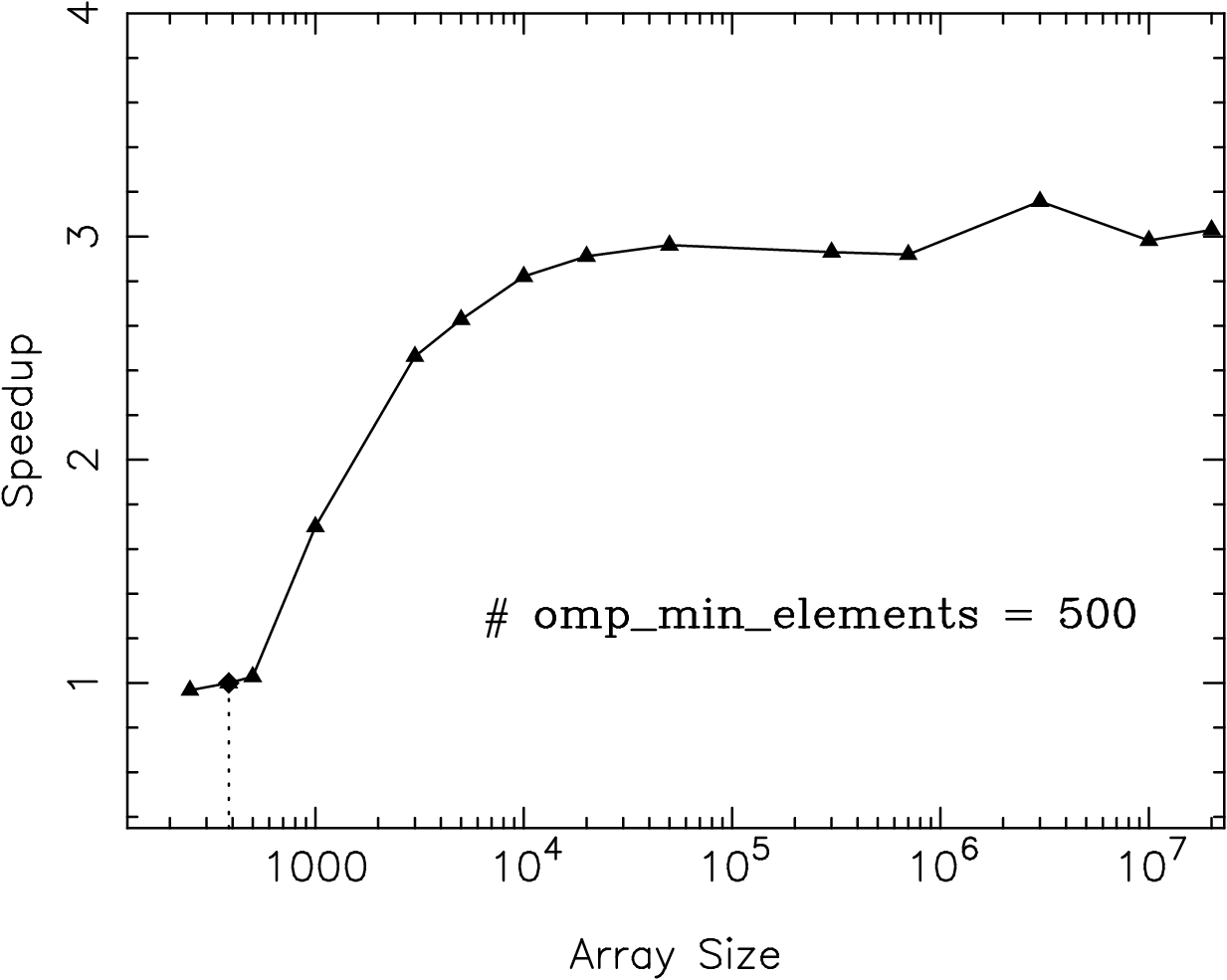}}}
  \caption{Aggregate speedup of the Weibull fit function due to the
  parallelized operators and functions detailed above.
  Left: \lint, with inflection point at 1907 array elements and mean
  runtimes from ca. 1.6e-4 to 23 sec. Right: \solf, with inflection point at
  384 elements and runtimes from ca 6e-4 to 61 sec.}
 \label{fit-perf}
\end{figure}

The speedup plots in Figs. \ref{funcs-perf}, \ref{opers-perf}, \&
\ref{fit-perf} demonstrate significant performance gains to be had
from parallelism.  Performance of the parallelized functions approaches
the theoretical maximum of linear speedup as array
sizes increase, and the inflection points in the size of the arrays needed
for nominal speedup from multithreading (represented by the dotted vertical
lines) are relatively small, ca. 1000 elements on \lint and 250 elements
on \solf. 

Fig.\ref{opers-perf}-(a) shows that on \lint the core arithmetic operators
did not parallelize as well as the functions, with speedups peaking at 60\%
before converging to 35\%, although the gains are respectable.  The lower
speedup is not surprising: these operators can be executed directly as
CPU instructions, requiring far less overhead than function invocation.
On \lint these instructions execute fast
enough to make the cost of parallel thread creation significant; on \solf
the same arithmetic instructions execute at a slower clock speed, resulting
in greater speedups from parallelism.  The relational and exponentiation
operators converged to nearly linear speedups on both platforms.  The
excellent speedup of exponentiation stems from the operator being implemented
in terms of the C {\tt pow()} function, so we should expect its speedup
curve to resemble those of Fig. \ref{funcs-perf}.  The relational
operators parallelized well because they are not atomic CPU operations;
they require many more assembly instructions to implement than, say, the
division operator, approaching the number required for a short function
call.
The large superlinear spikes in Figs. \ref{opers-perf} (b)-(d) appear
consistently in every dataset collected. They do not reflect faulty
parallelization of the \slang operators, because the serial and parallel
results were verified identical and similar trends were observed with
pure C codes written to perform the same computations.  We attribute
them to cache effects or page faulting that is less pronounced in parallel
execution because each CPU receives a smaller portion of the problem.

In Fig. \ref{fit-perf} these spikes are seen as smoother bumps at
the corresponding array sizes.  The Weibull model speedups converge
on ca. 50\% for \lint and 75\% for \solf.
While below the ideal of linear speedup, these are sizable
performance increases; models with more calls to parallelized
functions would exhibit even greater gains.  These results have
added significance in that end-users need to do nothing -- in terms
of learning parallelism or recoding sequential algorithms -- to obtain
them.
Furthermore,
recall that these models are used in the context of an iterative fitting
process.  Fits do not converge after just one iteration, and generating
accurate confidence intervals -- an absolute
necessity for credible modeling -- can require that thousands of fits
be performed at each point on a parameter space grid, with potentially
a million or more fits performed for a single pair of parameters, and
tens of millions if multiple parameter sets are to be explored.  In
such cases the speedups given here accumulate to significant differences
in the overall runtime of an analysis sequence.  By transparently using
\omp to effect greater multiprocessor utilization we gain the freedom
to explore on the desktop more challenging problems that other researchers
might avoid for their prohibitive cost of computation.

\section{Conclusion}
Multicore chip designs are making it possible for general users to access
many processors. At the granularity of the operating system it will be
relatively easy to make use of these extra cores, say by assigning whole
programs to separate CPUs.  As noted with increasing frequency of late,
though, it is not as straightforward to exploit this concurrency within
individual desktop applications.  In this paper we demonstrated how we
have helped our research colleagues prepare for this eventuality.
We have enhanced the vectorization capabilities of \slirp, a module
generator for the \slang numerical scripting language, so that wrappers
may be annotated for automatic parallelization with \omp.  This lets
\slang intrinsic functions be replaced with parallelized
versions, at runtime, without modifying a single line of internal \slang
source.  We have shown how \slang operators may also be parallelized with
relative ease, by identifying key loops within the interpreter source,
tagging them with \omp directives and recompiling.  These simple adaptations
have yielded beneficial speedups for computations actively used in
astrophysical research, and allow the same numerical scripts to be used
for both serial and parallel execution --
minimizing two traditional barriers to the use of parallelism by
non-specialists: learning how to program for concurrency and recasting
sequential algorithms in parallel form.  We extrapolate that the advent
of widespread \omp support in free compilers such as GCC presages a
proliferation of multicore-enabled scientific codes in the open source
community, parallelized in largely the manner given here.

\begin{acks}
This work was supported by NASA through the AISRP grant NNG06GE58G (HYDRA)
and Smithsonian Astrophysical Observatory contract SV3-73016 for the
Chandra X-Ray Center.  The author would like to thank his MIT colleagues
for thoughtful review and constructive criticism.
\end{acks}

\bibliographystyle{apalike}
\bibliography{multicore}

\begin{thebibliography}{}

\bibitem[{Beazley}, 1996]{1996SWIG}
{Beazley}, D. (1996).
\newblock {SWIG : An Easy to Use Tool for Integrating Scripting Languages with
  C and C++.}
\newblock In {\em 4th Tcl/Tk Workshop Proceedings}.

\bibitem[Chuang et~al., 2006]{1168901}
Chuang, W., Narayanasamy, S., Venkatesh, G., Sampson, J., Biesbrouck, M.~V.,
  Pokam, G., Calder, B., and Colavin, O. (2006).
\newblock {Unbounded page-based transactional memory}.
\newblock In {\em ASPLOS-XII: Proceedings of the 12th international conference
  on Architectural support for programming languages and operating systems},
  pages 347--358, New York, NY, USA. ACM Press.

\bibitem[Creeger, 2005]{1095423}
Creeger, M. (2005).
\newblock {Multicore CPUs For The Masses}.
\newblock {\em Queue}, 3(7):64--ff.

\bibitem[Creel, 2005]{1095677}
Creel, M. (2005).
\newblock {User-Friendly Parallel Computations with Econometric Examples}.
\newblock {\em Comput. Econ.}, 26(2):107--128.

\bibitem[{Folk} et~al., 1999]{1999IEEE...99...273F}
{Folk}, M., {McGrath}, R., and Yeager, N. (1999).
\newblock {HDF: an update and future directions}.
\newblock In {\em IEEE Geoscience and Remote Sensing Symposium, 1999, Vol. 1},
  pages 273--275.

\bibitem[{Houck}, 2002]{2002hrxs.confE..17H}
{Houck}, J.~C. (2002).
\newblock {ISIS: The Interactive Spectral Interpretation System}.
\newblock In {Branduardi-Raymont}, G., editor, {\em High Resolution X-ray
  Spectroscopy with XMM-Newton and Chandra}.

\bibitem[Karlsson and Brorsson, 2004]{OdinMP.Nov2004}
Karlsson, S. and Brorsson, M. (2004).
\newblock {A Free OpenMP Compiler and Run-Time Library Infrastructure for
  Research on Shared Memory Parallel Computing}.
\newblock In {\em Proceedings of The 16th IASTED International Conference on
  Parallel and Distributed Computing and Systems}, Calgary, AB, Canada. Acta
  Press.

\bibitem[Kusano et~al., 2000]{690388}
Kusano, K., Satoh, S., and Sato, M. (2000).
\newblock {Performance Evaluation of the Omni OpenMP Compiler}.
\newblock In {\em ISHPC '00: Proceedings of the Third International Symposium
  on High Performance Computing}, pages 403--414, London, UK. Springer-Verlag.

\bibitem[Lee, 2006]{10.1109/MC.2006.180}
Lee, E.~A. (2006).
\newblock {The Problem with Threads}.
\newblock {\em Computer}, 39(5):33--42.

\bibitem[Massaioli et~al., 2005]{1131942}
Massaioli, F., Castiglione, F., and Bernaschi, M. (2005).
\newblock {OpenMP Parallelization of Agent-Based Models}.
\newblock {\em Parallel Comput.}, 31(10-12):1066--1081.

\bibitem[{Noble} et~al., 2006]{2006ASPC..351..481N}
{Noble}, M.~S., {Houck}, J.~C., {Davis}, J.~E., {Young}, A., and {Nowak}, M.
  (2006).
\newblock {Using the Parallel Virtual Machine for Everyday Analysis}.
\newblock In {Gabriel}, C., {Arviset}, C., {Ponz}, D., and {Enrique}, S.,
  editors, {\em ASP Conf. Ser. 351: Astronomical Data Analysis Software and
  Systems XV}, pages 481--+.

\bibitem[Weibull, 1939]{Weibull.1939}
Weibull, W. (1939).
\newblock {A Statistical Theory of the Strength of Materials}.
\newblock {\em Ingenior Ventenskaps Akademien Handlinger}, 151.

\end{thebibliography}

\end{document}